\documentclass{aa}
\usepackage{graphicx}
\usepackage{txfonts}

\begin{document}
   \title{Optical data of meteoritic nano-diamonds from far-ultraviolet to 
   far-infrared wavelengths\footnotemark[1]}

%   \subtitle{}

   \author{Harald Mutschke\inst{1}
          \and
          Anja C.\,Andersen\inst{2}
          \and
          Cornelia J\"ager\inst{1}
          \and
          Thomas Henning\inst{3}
	  \and
          Anja Braatz\inst{1}
          }

   \offprints{H. Mutschke}

   \institute{
     Astrophysikalisches Institut und Universit\"ats-Sternwarte
     (AIU), Schillerg\"a\ss{}chen 3, D-07745 Jena, Germany \\
   \email{mutschke@astro.uni-jena.de, conny@astro.uni-jena.de}
   \and
    NORDITA, Blegdamsvej 17, DK-2100 Copenhagen, Denmark \\
   \email{anja@nordita.dk} 
   \and
   Max-Planck-Institut f\"ur Astronomie, K\"onigstuhl 17, 
   D-69117 Heidelberg, Germany\\
   \email{henning@mpia.de}
     } 
   \date{Received 21 October 2003; accepted 10 May 2004}

   \abstract{We have used different spectroscopic techniques to obtain a 
consistent quantitative absorption spectrum of a sample of meteoritic 
nano-diamonds in the wavelength range from the vacuum ultraviolet 
(0.12 $\mu$m) to the far infrared (100 $\mu$m). The nano-diamonds have been 
isolated by a chemical treatment from the Allende meteorite (Braatz 
et al.\,2000). Electron energy loss spectroscopy (EELS) extends the 
optical measurements to higher energies and allows the derivation of 
the optical constants (n\&k) by Kramers-Kronig analysis. 
The results can be used to restrain observations and to improve current 
models of the environment where the nano-diamonds are expected to have formed. 
We also show that the amount of nano-diamond which can be present in space 
is higher than previously estimated by Lewis et al.\,(1989). 
    }
   \authorrunning{Mutschke et al.} 
   \titlerunning{Optical data of meteoritic nano-diamonds}
   \maketitle
%
%________________________________________________________________

\section{Introduction}

\footnotetext[1]{The n and k data in Table 3 are also available 
in electronic form at the CDS via anonymous ftp to cdsarc.u-strasbg.fr 
(130.79.128.5) or via http://cdsweb.u-strasbg.fr/cgi-bin/qcat?J/A+A/ 
as well as from http://www.astro.uni-jena.de//Laboratory/Database/carbon.html}

Presolar meteoritic grains have opened the possibility of studying circumstellar 
dust grains directly in the laboratory, providing an important complement to 
astronomical observations. The identification of the presolar origin of the 
grains is based on their highly anomalous isotopic composition which in general 
agrees with those expected for stellar condensates (Zinner \cite{zinner98}; Ott \cite{ott03}). 
Diamonds were the first presolar grains to be isolated from meteorites 
(Lewis et al.\ \cite{lewis+etal87}) and they account for more 
than 99\% of the identified presolar meteoritic material. Their abundance 
normalized to the meteorite's matrix is around 500$-$1000 ppm, depending on 
the metamorphic grade of the meteorite (Alexander et al.\ \cite{alexander+etal90}; 
Huss \cite{huss90}; Huss \& Lewis \cite{huss+lewis95}). They are identified 
as presolar grains due to a significant overabundance (compared to the Solar 
isotopic ratios) of the very heavy (Xe-H) $^{134}$Xe and $^{136}$Xe as well as the very 
light (Xe-L) $^{124}$Xe, $^{126}$Xe isotopes (together called Xe-HL composition). 

The origin of the meteoritic nano-diamonds is still a matter of debate. The Xe-HL anomaly 
is expected to be produced by $r$- and $p$-process nucleosynthesis in supernovae (SN). 
Therefore, it has been suggested that the diamonds condense either directly in 
SN outflows (Clayton \cite{clayton89}) or that they originate from 
cool red giants and are later implanted with Xe (J{\o}rgensen \cite{jorgensen88}). 
As proposed by J{\o}rgensen \& Andersen (\cite{jorgensen+andersen99}) it is 
also possible that the Xe-H anomaly can be explained as a small fraction of pure 
$^{12}$C diamonds which originated in SN~{\small II}. The bulk of the presolar 
meteoritic diamonds could then have condensed in evolved carbon stars which 
would account for the bulk $^{12}$C/$^{13}$C ratio. The content 
of light isotopes of Xe-L is very small (less than 1 atom per 10$^9$ diamonds), 
and can be explained as coming from SN~{\small I} (Lambert \cite{lambert92})
in binary systems where the low mass component is an evolved carbon star
(J{\o}rgensen \cite{jorgensen88}), or as a by-process of the
Xe production in SN~{\small II} (Ott \cite{ott96}). 
However, recently Dai et al.\ (\cite{dai+etal02}) have, based 
on investigations of interstellar dust particles, suggested that at least 
some of the meteoritic nano-diamonds were formed in the inner solar nebula 
and not in a presolar environment. Whether this would be consistent 
with the oxidizing conditions expected for the solar nebula is an open question. 

So far, the only strong observational evidence for the presence of diamonds in 
other astrophysical environments comes from the identification of the infrared 
emission bands of hydrogen-saturated diamond surfaces at 3.43 and 3.53 $\mu$m 
in certain Herbig-AeBe objects as well as in the post-AGB star HR~4049 
(Guillois et al.\ \cite{guillois+etal99}; van Kerckhoven et al.\ 
\cite{vankerckhoven+etal02}). 
The earlier identification by Allamandola et al.\ (\cite{allamandola+etal92}), 
who suggested nano-diamonds as the source of a 3.47 $\mu$m absorption feature 
seen in dense molecular clouds (protostars), now seems untenable (Brooke et al. \cite{brooke+etal96}; 
\cite{brooke+etal99}). The $\sim 21 \mu$m feature seen in some 
post-AGB stars (Kwok et al.\ \cite{kwok+etal89}) is attributed by Hill et al.\ 
(\cite{hill+etal98}) to the presence of nano-diamonds, however this feature has
also been attributed to several other molecules and solids (for recent 
assignments see Kwok et al.\ \cite{kwok+etal99}; Volk et al.\ \cite{volk+etal99}; 
Papoular \cite{papoular00}; von Helden et al.\ \cite{vonhelden+etal00}; 
Speck \& Hofmeister \cite{speck+hofm03}, Posch et al.\ \cite{posch+etal04}).

The diamond grains seen in the Herbig-AeBe objects 
are different from meteoritic ones. The observed emission 
bands require that the diamonds must be at least one order of magnitude 
larger than the average size of the meteoritic grains. Nano-diamonds 
of the size of the meteoritic ones would emit broader bands at about 
3.41 and 3.50~$\mu$m (Sheu et al.\, \cite{sheu+etal02}; 
Jones et al.\ \cite{jones+etal04}), provided that their surfaces 
are hydrogenated. Jones et al.\ \cite{jones+etal04} have recently 
published a new study on the C-H stretching IR bands of surface-processed 
meteoritic nano-diamonds. They find that the spectra of nano-diamonds 
of a certain surface structure show similarities with class B 
emission band sources. They also emphasize that this would strengthen 
the case for a nano-diamond carrier of the $\sim 21 \mu$m emission feature 
observed in most of the class B objects. Altogether, the current 
observational and experimental findings, including the presence of 
the meteoritic nano-diamonds indicate, that diamonds are a 
constituent at least of certain astrophysical environments. 
Nano-diamonds with sizes like the meteoritic diamonds could very 
well be more common than the larger diamonds, due to their detection 
limitations.

If nano-diamonds are present, they need to be taken into account in models 
of these environments. Especially, the optical properties (absorption and scattering 
cross sections) will have to be included in modelling of synthetic spectra. 
This requires knowledge of the optical constants of the material within a 
broad wavelength range, i.e.\ the complex refractive index ($m = n + k$) 
or the dielectric function ($\varepsilon = n^2$).
These data have so far been available only in limited spectral ranges. Lewis 
et al.\ (\cite{lewis+etal89}) made the attempt to combine measured infrared data 
with modified literature data of bulk diamonds in the ultraviolet. 
Part of the UV/Vis and IR spectrum of meteoritic nano-diamonds was 
measured by  Mutschke et al.\ (\cite{mutschke+etal95}), Andersen et al.\ 
(\cite{andersen+etal98}), and Braatz et al.\ (\cite{braatz+etal00}).

With this paper we intend to provide the needed optical data by combining 
information from optical and EELS measurements on a sample of highly 
purified meteoritic nano-diamonds extracted from the Allende meteorite 
(Braatz et al.\ \cite{braatz+etal00}). After a short summary of the physical and 
chemical characteristics of the meteoritic nano-diamonds, which are essential 
to understand their spectroscopic properties, the optical measurements 
are described in Sect.~3,  followed by the EELS measurements and the 
derivation of the complex refractive index (Sect.~4). 
In Sect.~5 we discuss the consequences of our new results for the 
observability of nano-diamonds in astrophysical environments. 

\section{Properties of the meteoritic nano-diamonds}

\begin{figure}
\centering\includegraphics[width=8cm,angle=0,clip]{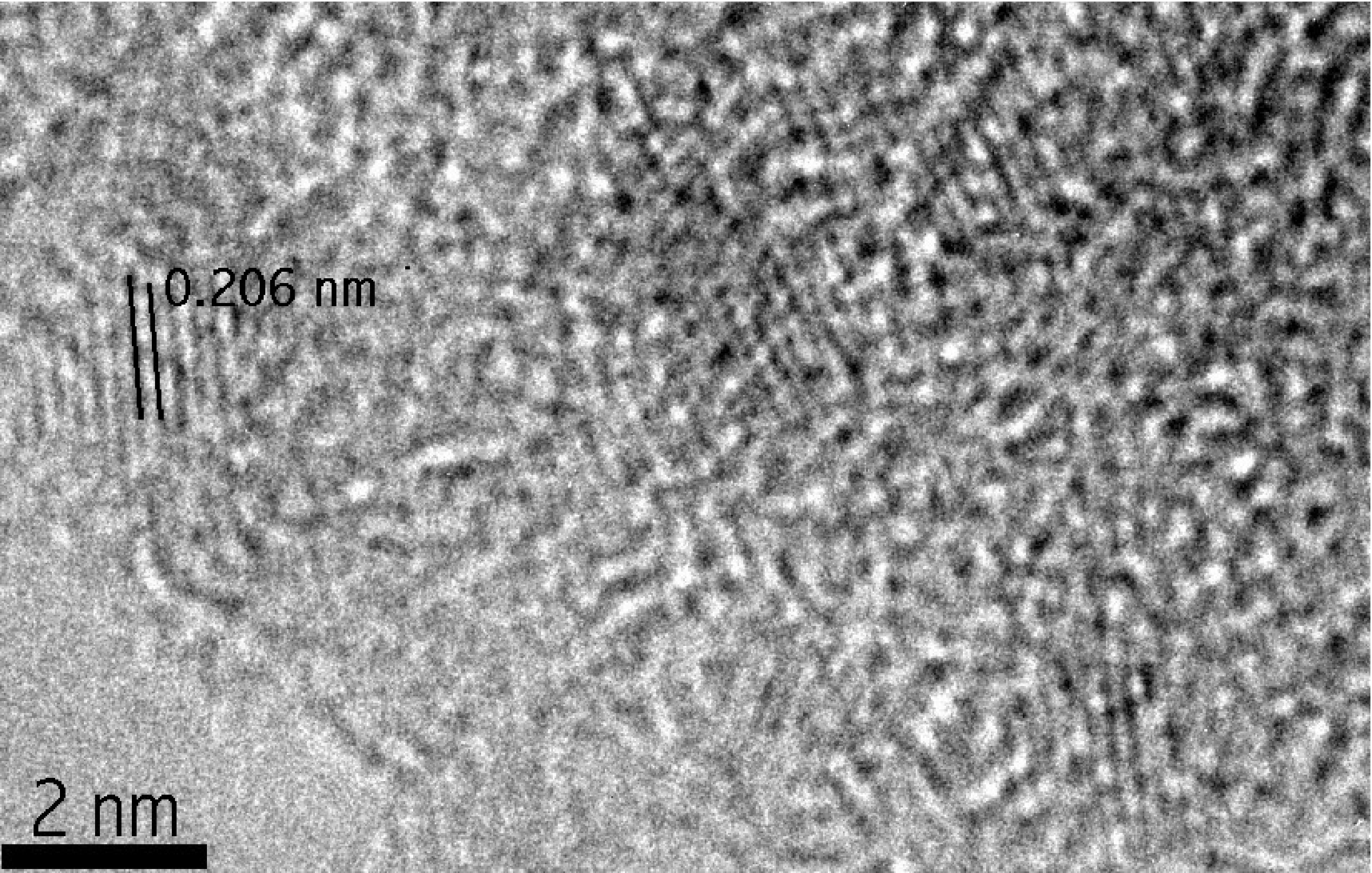}
\centering\includegraphics[width=8cm,angle=0,clip]{0544f1b.eps}
\caption{{\it Upper panel}: bright-field transmission electron micrograph 
showing an agglomerate of meteoritic nano-diamonds from our sample 
with characteristic fringes corresponding to the (111) spacing of 
the diamond lattice. {\it Lower panel}: EELS core-loss spectrum of the 
meteoritic nano-diamonds showing the 1s-$\sigma^\star$ transition 
bands of diamond (B--D). A denotes the position of contributions 
from 1s-$\pi^\star$ transitions.}
\label{HRTEM}
\end{figure}

Presolar diamonds are found in relatively unprocessed meteorites, 
the so-called carbonaceous chondrites, and they account for about 
3\% of the total amount of carbon in the meteorite. 
An in-situ search for diamonds in the Allende meteorite (Banhart et al.\ 
\cite{banhart+etal98}) showed that they are situated in the matrix of 
the meteorite, i.e. not related to other inclusions such as the chondrules. 
A number of impurities have been identified in the meteoritic 
diamonds, including noble gases (He, Ne, Ar, Kr, and Xe), Ba and Sr
(which are slightly enriched in $r$-process isotopes; Lewis et al.\ 
\cite{lewis+etal91}), H with $^1$H/$^2$D = 5193 (Virag et al.\ 
\cite{virag+etal89}; ($^1$H/$^2$D)$_{\rm terrestrial}$ = 6667),
N with $^{14}$N/$^{15}$N = 406 (Russell et al.\ \cite{russell+etal91};
($^{14}$N/$^{15}$N)$_{\rm terrestrial}$ = 272). The $^{12}$C/$^{13}$C ratio 
is only slightly higher than the terrestrial one (92 compared to 89; 
Lewis et al.\ \cite{lewis+etal87}). 

The most intriguing property of the nano-diamonds is their very small 
size, with a median diameter of less than 20\,\AA\ 
(Fraundorf et al.\ \cite{fraundorf+etal89}) and, consequently, 
the dominance of the grain surface structure in their physical and 
chemical properties. A typical meteoritic diamond contains on the order 
of 1000 - 2000 carbon atoms, with $\approx$ 25\% of these belonging to the 
surface (Jones \& d'Hendecourt \cite{jones+dhendecourt00}). 
Since the surface atoms have free valences, they can either bind chemical 
groups or, in the case of carbon atoms, change their hybridization state 
and restructure the C-C bonds, i.e., form unsaturated bonds (for 
detailed discussions on nano-diamond surface structure see Evans 
\cite{evans92}; Jones \& d'Hendecourt \cite{jones+dhendecourt00} 
and references therein).

Indications for this restructuring of the surface come from electron 
energy loss spectroscopy (EELS) as well as from density measurements. 
Bernatowicz et al.\ (\cite{bernatowicz+etal90}) found that in EELS spectra 
of nano-diamonds extracted from the Allende and Murray CE meteorites 
the diamond ($\sigma$-electron) plasmon was shifted to smaller energies 
by about 5~eV (see also Sect.~4.1). They explained this behaviour by 
assuming a grain mantle consisting of hydrogenated amorphous carbon and 
estimated the volume fraction of this mantle to be 0.46 of the grain volume. 
Lewis et al. (\cite{lewis+etal87}) have measured an effective bulk 
density of meteoritic nano-diamonds from the Murchison meteorite of 
2.22-2.33~g~cm$^{-3}$, well below the density of pure diamond 
(3.51 g~cm$^{-3}$). Lewis et al.\, (\cite{lewis+etal89}) interpret 
this as an implication for the presence of residual water and 
amorphous carbonaceous phases (a-C and a-C:H), which was taken 
by Bernatowicz et al.\ (\cite{bernatowicz+etal90}) as support for 
the core-mantle model, which would give a density of 2.4-2.5~g~cm$^{-3}$.

Another clear indication for sp$^2$ carbon is the presence of a 
1s-$\pi^\star$ feature in the carbon core-electron transitions 
in EELS spectra measured by several authors (Bernatowicz et al.\ 
\cite{bernatowicz+etal90}; Blake et al.\ \cite{blake+etal88}; 
Dorschner et al.\ \cite{dorschner+etal96}, 
see also Fig.\, \ref{HRTEM}). The analysis of the strength of the 
feature in the present sample gives a typical fraction of sp$^2$/sp$^3$ 
carbon atoms of 0.11, i.e. about half of the surface atoms are sp$^2$. 
This would be consistent with the presence of C=C chains at the 
restructured surface, which has been described by Evans (\cite{evans94}), 
and with a contribution from C=O groups. The abundance of C=O groups 
can be estimated from the IR spectrum (5.7~$\mu$m band) to be about 
1/4 of the sp$^2$ atoms. The sp$^3$ surface atoms probably 
largely form C-O-C and C-OH groups. 

The importance of the surface properties is underlined by infrared spectroscopy 
results, which reveal a number of bands due to functional groups bound to the 
nano-diamond surface. These bands have been measured before by several groups on 
nano-diamonds from different meteorites such as Allende (Lewis et al.\ \cite{lewis+etal89}; 
Koike et al.\ \cite{koike+etal95}; Andersen et al.\ \cite{andersen+etal98}; 
Braatz et al.\ \cite{braatz+etal00}), Murchison (Colangeli et al.\ \cite{colangeli+etal94}; 
Mutschke et al.\ \cite{mutschke+etal95}; Braatz et al.\ \cite{braatz+etal00}) 
and Orgueil (Hill et al.\ \cite{hill+etal97}). Comparisons of the measurements 
can be found in Andersen et al.\ (\cite{andersen+etal98}) and Braatz et al.\ 
(\cite{braatz+etal00}). The assigned bands indicate the presence of oxygen, 
hydrogen, and nitrogen in different configurations at the diamond surface. 

As mentioned before, the results of the isotopic analysis 
(Russell et al.\, \cite{russell+etal91}; Virag et al.\ \cite{virag+etal89}) 
imply that at least some of the H and N must be presolar. 
However, by IR spectroscopy it has not yet been possible to identify 
surface groups which originate from the astrophysical environment 
in which the grains have either been formed or processed. Before it is 
possible to measure the spectral properties of meteoritic diamonds, 
it is necessary to extract them from the rest of the meteoritic material. 
This means that 99.9\% of the meteorite has to be removed in a destructive 
chemical separation, in which undesirable minerals are dissolved by 
appropriate reagents. Hill et al. (\cite{hill+etal97}) and 
Braatz et al. (\cite{braatz+etal00}) showed that e.g. the oxidizing steps 
of the extraction procedure should be able to modify the surface structure. 

The chemical separation procedure we have used, is a variant of those 
developed by Tang \& Anders (\cite{tang+anders88}) and Amari et al.\, 
(\cite{amari+etal94}) and is described in Braatz et al.\ (\cite{braatz+etal00}) 
together with infrared spectra taken after each step for a sample of Murchison 
diamonds. The nano-diamond sample used for the measurements presented here 
is identical to the Allende sample of Braatz et al.\ (\cite{braatz+etal00}). 
The diamonds were extracted from 114~g of the Allende meteorite. 
The extraction procedure is the same for these. The IR spectra are reported 
in Braatz (\cite{braatzthesis}). The yield of the final diamond 
sample was 16 mg.

\section{Optical spectroscopy}

We have performed optical transmission measurements of meteoritic nano-diamonds 
in the wavelength range $0.12 - 100~\mu$m, which required the use of different 
spectrometers and preparation methods. To prepare suitable samples, the 
diamonds had to be either dispersed in or deposited on media which are 
transparent in the corresponding part of the spectral range. Because of the 
strong variation in the mass absorption coefficient from the vacuum-ultraviolet 
to the far-infrared wavelengths, samples with different column densities 
of nano-diamonds had to be prepared.  The intended quantitative absorption 
measurements require as exact a knowledge of the column densities as possible. 

Dispersion in liquids failed, because the diamonds exhibited a strong tendency 
to aggregate and to sediment, regardless of the pH-value or kind of liquid. 
We therefore used the following methods: (1) 
the sample was dispersed in KBr and polyethylene pellets for measurements 
in the visible and the infrared spectral ranges and (2) the sample was 
deposited onto CaF$_2$ windows from a suspension for measurements in the UV. 
Table~\ref{methods} summarizes the methods and instruments applied for each 
of the different spectral regions. 

\begin{table*}
\caption{Measuring methods for spectroscopy of the meteoritic nano-diamonds in 
different parts of the spectral region.}
\begin{center}
\begin{tabular}{|c|c|c|c|c|c|} \hline
Wavelength & Preparation & Column density & Instrument & Resolution & Remarks \\
($\mu$m) & &[mg/cm$^2$] & & $\lambda$/$\Delta\lambda$ & \\ \hline
0.01 - 0.2  & on TEM grid & - & EELS (GIF 100) & 200 - 10 & -  \\ \hline
0.12 - 0.23 & on CaF$_2$ substrate & 0.0169 & VUV (LZ Hannover) & $\sim$ 500 & closed film \\ \hline
0.19 - 0.5 & on CaF$_2$ substrate & 0.069 & Lambda 19 (Perkin Elmer) & $\sim$ 3000 & integrating sphere \\ \hline
0.5 - 2 & in KBr pellet & 1.81 & Lambda 19 (Perkin Elmer) & $\sim$ 10000 & agglomeration \\ \hline
2 - 25 & in KBr pellet & 0.52 & FTIR 113v (Bruker) & 2500 - 200 & agglomeration \\ \hline
25 - 100 & in PE pellet & 1.57 & FTIR 113v (Bruker) & 400 - 100 & agglomeration \\ \hline
\end{tabular}
\end{center}
\label{methods}
\end{table*}

\subsection{Measurements on films on CaF$_2$ substrate}

For the vacuum-ultraviolet (VUV) and the ultraviolet/visible (UV/Vis) measurements, 
four diamond-film samples were produced by transferring 3.7, 8.3, 34.4, and 
63.3\,$\mu$g, respectively, of nano-diamonds from an aqueous colloidal 
(pH=6.8) to the surface of CaF$_2$ windows. After drying, the nano-diamonds 
formed closed films of 7-8~mm diameter. The covered area was measured under 
an optical microscope. The deposited mass was determined with an accuracy 
of $\pm$0.05\,$\mu$g by weighing of the windows before and after deposition 
of the diamonds. The thickness homogeneity of the films was checked by optical 
transmission measurements at different positions and was found to be better 
than $\pm$20~\%. 

In the UV and visible (190 - 800~nm), a Perkin Elmer Lambda 19 spectrometer 
was used for measuring the transmission through the diamond films. The VUV 
spectrometer was a double-beam grating spectrometer built by Laserzentrum Hannover, 
using a deuterium lamp with MgF$_2$ window as the light source. The wavelength range 
of this spectrometer is 115 - 230~nm (accuracy 0.2~nm), but the range of the 
measurements was limited by the CaF$_2$ substrates to $\ge$124~nm. The VUV and 
UV/Vis transmission spectra measured on the same sample always agreed within a 
relative error of $\pm$5~\% in the overlap region. 

In order to eliminate substrate influences, in the VUV the transmission was 
measured at each of the CaF$_2$ windows before and after sample deposition. 
The ratio of these spectra is the transmittance of the diamond film, which 
contains information about the reflection, the scattering and the absorption 
by the film. With the Lambda 19 spectrometer, we are able to disentangle these 
different loss mechanisms by additional measurements using an integrating sphere 
detector (spectral range 0.25 - 2.0~$\mu$m). For these measurements, the 
sample is placed inside the integrating 
sphere and the intensity coming from the illuminated sample is collected 
and compared to the intensity of a reference beam. Opening/closing of ports 
for the transmitted and specularly reflected beams makes it possible to measure 
the scattered light individually or together with these contributions. 
Closing of all ports allows a direct measurement of the intensity absorbed by 
the sample (``absorption mode''). 

\begin{figure}
\includegraphics[width=8.5cm,angle=0,clip]{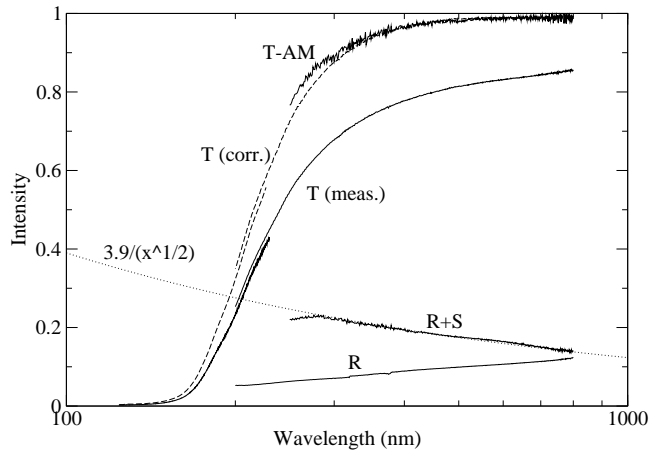}
\caption{Illustration of the scattering correction for the VUV and UV/Vis 
transmission spectra T(meas.) measured at the 69~$\mu$g~cm$^{-2}$ diamond film. 
The solid lines represent the original measurements, where the ``reflectance 
plus scattering'' (R+S), and ``absorption mode'' (T-AM) spectra have been measured 
using an integrating sphere. The dotted line is a fit to the R+S spectrum, 
which has been used to produce the corrected spectra (T/(1-R-S), dashed lines). 
The corrected UV/Vis spectrum corresponds well to the ``absorption mode'' 
integrating sphere measurement. }
\label{integratingsphere}
\end{figure}

The measurements using the integrating sphere revealed a reflectance of the 
samples of the order of 13~\% at $\lambda$ = 800~nm, which is expected for a 
film with a refractive index of about 2 on CaF$_2$ substrate 
(see Fig.\,\ref{integratingsphere}). This value decreased with decreasing 
wavelength, accompanied by an increase in the intensity of the diffusely 
scattered light. For the thinner films, scattering plus reflection increased 
only slightly toward smaller wavelengths, whereas for thicker films it 
increased up to 22~\% at $\lambda$ = 300~nm. We interpret this as a higher 
surface roughness of the thicker films. 

The correction of the transmission spectra for the scattering and reflection 
effects was done by taking the reduction of the incident intensity by reflection 
and scattering losses into account, i.e. by dividing the original spectra 
by 1--R--S, R being the reflectance and S the scattering. For this purpose, 
the R+S spectra have been fitted either by a 
linear or a $\lambda^{-1/2}$ function and extrapolated to the VUV wavelengths. 
The resulting corrected spectra agree well with the spectra measured in the 
``absorption mode'' (see Fig.\,\ref{integratingsphere}). However, because 
of the narrower spectral region and the stronger noise, the latter have not been used. 

For the final spectrum, in the VUV (wavelength range 0.12 - 0.23~$\mu$m), 
the spectrum measured at the 8.3~$\mu$g sample (column density $\sigma$ = 16.9~$\mu$g~cm$^{-2}$) 
was selected, because this was the thickest film being still transparent 
at the shortest wavelengths. For the UV/Vis wavelengths up to 0.5~$\mu$m, 
the spectrum of the 34.4~$\mu$g sample (column density $\sigma$ = 69~$\mu$g~cm$^{-2}$) 
was used. For significant measurements at larger wavelengths, the 
absorption by these films was too low. The thickest film (63.3~$\mu$g) 
exhibited a tendency to flake and was not used. 

The transmission T was converted to the mass absorption coefficient 
$\kappa_{film} \equiv \alpha / \rho = \ln (1/T) / \sigma$. Here, $\rho$ is 
the mass density and $\alpha$ is the volume absorption coefficient which 
for a continuous material is given by $\alpha = 4 \pi k / \lambda$ 
with $\lambda$ being the wavelength and $k$ being the 
imaginary part of the complex refractive index. 

\subsection{Measurements on KBr and polyethylene pellets}

The spectrum in the visible and near-infrared range up to a wavelength of 
2 $\mu$m were measured again with the Lambda 19 spectrometer equipped with 
the integrating sphere. 
The sample here consisted of 2.4~mg of nano-diamonds, mixed thoroughly with 
powdered KBr and pressed into a pellet of 13~mm diameter. The column 
density of the nano-diamonds was thus $\sigma$ = 1.81~mg~cm$^{-2}$. The pellet 
showed very strong diffuse scattering of up to 50\% of the incident 
intensity, probably caused by big diamond aggregates concentrated around 
the KBr grains and by the high refractive index of the nano-diamonds 
compared to the KBr matrix. Consequently, apart from measuring the scattered 
intensity S, the ``absorption mode'' (see Sect. 3.1) of the integrating 
sphere was used to determine the absorbed intensity A. Since the scattering 
here is a volume effect in contrast to the scattering from the diamond film, 
scattering and absorption have to be treated as directly competing mechanisms, 
and the mass absorption coefficient can be calculated from 
\begin{equation}
\kappa_{\rm abs} \, = \, - \, \frac{1}{\sigma} \, {\rm ln} \, (1 - A - S) \, \times \, \frac{A}{A + S}.
\end{equation}
The calculated mass absorption coefficient is used in the final spectrum 
in the wavelength range 0.5 - 2~$\mu$m. At shorter wavelengths, it 
starts to fall below the values expected from the film measurements because 
of saturation effects in the big clusters. 

The measurements with the integrating sphere demonstrate that at infrared 
wavelengths nearly all the continuum extinction produced by nano-diamonds 
in KBr pellets is due to light scattering. At a wavelength of 2~$\mu$m, 
$\kappa_{\rm abs}$ has fallen to about 20~cm$^2$g$^{-1}$, whereas the 
extinction measured with the pellets is typically many 100~cm$^2$g$^{-1}$, 
even for pellets with a lower diamond concentration 
(see Braatz et al.\, \cite{braatz+etal00}). This was also noticed by 
Andersen et al.\, (\cite{andersen+etal98}), but a quantitative determination 
of the continuum absorption coefficient requires the integrating sphere 
technique and was not possible at that time. 

Since the use of the integrating sphere is limited to wavelengths below 
2~$\mu$m, at larger wavelengths quantitative continuum absorption 
measurements with the KBr pellet technique have not been possible. 
However, with other techniques we did not succeed in producing the 
diamond column density of $\sim$ 10~mg cm$^{-2}$ required for continuum 
measurements at mid-IR wavelengths. Therefore, for extending the diamond 
spectrum to the mid-IR, we have corrected the available spectra from 
Braatz (\cite{braatzthesis}) for the scattering background and fitted 
them to the near-IR spectrum at $\lambda$ = 2~$\mu$m assuming that the 
slope of the spectrum between 1 and 2~$\mu$m, which is approximately 
proportional to $\lambda^{-1}$, would continue throughout the infrared. 
This slope is justified for nano-grains with disordered surfaces 
(Seki \& Yamamoto \cite{sekiyamamoto80}), and agrees also well 
with the results of simple model calculations taking the disordered 
surface into account (see Fig.\, \ref{mac}). 

\subsection{Description of the final spectrum}

\begin{table}
\caption{Assigned absorption bands in the infrared spectra 
of the diamonds from the Allende meteorite. }
\begin{center}
\begin{tabular}{|c|c|l|} \hline
$\lambda$ & $1/\lambda$ & vibrational groups \\
($\mu$m) & (cm$^{-1}$) & \\ \hline
2.95 & 3400 & OH stretching \\
5.70 & 1760 & C==O stretching (Ester) \\
6.2 & 1610 & OH deformation, \\
  & & C==C stretching \\
7--11 & 1400--900& C--O--C (Ether) \\
  & & C--O(H) \\ 
  & & C==C deformation\\
  & & C--C stretching\\ 
19.7 & 508 & chromite \\ 
26 & 385 & OH \\ \hline
\end{tabular}
\label{surfacegroups}
\end{center}
\end{table}

\begin{figure*}
\includegraphics[width=15cm,angle=0,clip]{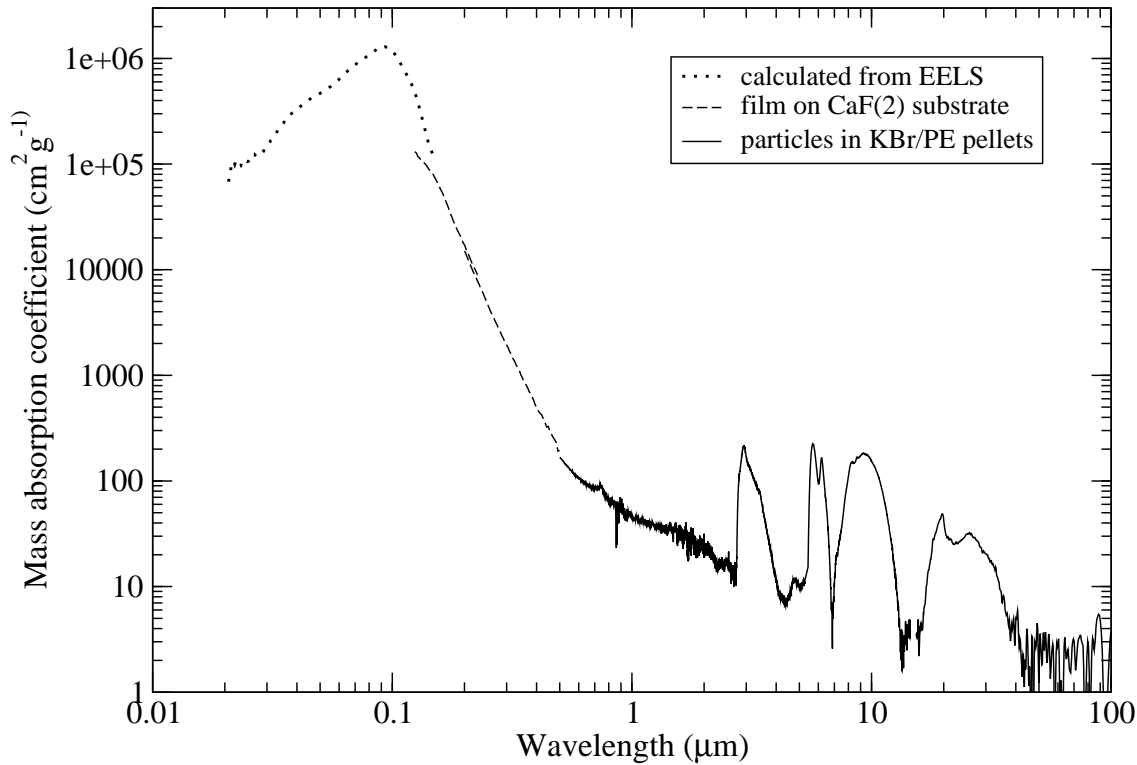}
\caption{The combined optical spectra are characterized by a smooth curve 
throughout the UV, visible, and near infrared. In the IR absorption bands
appear due to the excitation of surface groups (see Table\,\ref{surfacegroups}
for the identification).
The ``artificial'' far-UV spectrum, calculated from the EELS data, shows the 
inter-band electronic transitions produced by the carbon $\sigma$ electrons. 
At larger wavelengths, there is no evidence for $\pi$ electronic transitions 
or bands caused by nitrogen defect centers.}
\label{fullspec}
\end{figure*}

The combined optical absorption spectrum is characterized by a smooth curve 
throughout the UV, visible and near IR. In the UV at a wavelength of 200~nm, 
the mass absorption coefficient reaches about 10$^4$~cm$^2$g$^{-1}$, which is 
at least one order of magnitude less than for amorphous carbon but higher than 
for terrestrial diamonds. It is also higher by about a factor of two than 
the values reported by previous studies (Mutschke et 
al.\, \cite{mutschke+etal95}; Andersen et al.\, \cite{andersen+etal98}; 
Braatz et al.\, \cite{braatz+etal00}) for nano-diamonds dispersed in liquids. 
An explanation for this difference could be the sedimentation of diamond 
aggregates from the liquid lowering the actual concentration and therefore the 
measured absorption. Another explanation could be the influence of the 
thin-film geometry. This will be discussed further below. 

There is no evidence for any of the features in the $200-300$~nm range, which 
were reported by Mutschke et al.\ (\cite{mutschke+etal95}), Andersen et al.\
(\cite{andersen+etal98}) and Braatz et al.\ (\cite{braatz+etal00}) 
with meteoritic diamonds. Whereas the feature at about 220~nm has only been 
seen by Mutschke et al.\ (\cite{mutschke+etal95}) and is probably due 
to a contamination, the 270~nm feature was detected by all authors and was 
ascribed by Braatz et al.\ (\cite{braatz+etal00}) to singly distributed N atoms. 
However, Andersen et al.\ (\cite{andersen+etal98}) noted that the feature 
was much weaker in Allende than in Murchison nano-diamonds, which is consistent 
with the lower N concentration (Russell et al.\ \cite{russell+etal91}). 
In previous measurements on the same sample as studied here, but dispersed 
in liquid, one of us found a very weak indication of enhanced absorption 
around 300~nm (Braatz \cite{braatzthesis}). 

It is possible that this weak band is also present in our result but that 
it is too weak to be detected on the background of the absorption edge. 
The difference in the spectra published by Andersen et al.\ (\cite{andersen+etal98}), 
and Braatz et al.\ (\cite{braatz+etal00}) and this study, however, leaves the 
possibility open that this band might be due to contamination as well. Graphitic 
carbon, e.g., would be able to cause a band at a similar position. 
For the purpose of this paper, this question is not very important because the 
influence of the band on the broad-band optical properties is only marginal. 

In the IR, there appear absorption bands due to excitation of surface groups. 
The positions of the bands as well as their assignment are summarized in 
Table~\ref{surfacegroups}. They are clearly dominated by oxygen in 
carboxyl, carbonyl, and C-O-C groups. The influence of the isolation 
procedure on these bands is discussed in detail by 
Mutschke et al.\, (\cite{mutschke+etal95}), Hill et al.\ \cite{hill+etal97} 
and Braatz et al.\ (\cite{braatz+etal00}). This will not be repeated here. 

It is an open question as to which of the features could have 
originated from the presolar history of the diamonds. 
There is an ongoing discussion on the possible formation sites of 
nano-diamonds; carbon-rich AGB stars (J{\o}rgensen \& Andersen 
\cite{jorgensen+andersen99}), supernovae (Clayton \cite{clayton89}, 
Ott \cite{ott96}), the ISM (Tielens et al.\, \cite{tielens+etal87}) 
and the Solar nebula (Dai et al.\, \cite{dai+etal02}) have been 
suggested. The surface features of the nano-diamonds should first reflect 
the chemical environment at their place of formation. In carbon-rich 
AGB stars the dangling bonds of the diamond surface should be expected 
to primarily be chemically saturated by  hydrogen and deuterium. 
In contrast to that, the other formation places are oxygen-rich 
which will lead to the formation of oxygen-containing functional groups 
at the surface as well as the graphitization of the surface due to 
catalytic reactions (Evans \cite{evans92}). 

Furthermore, the surface structure is likely to be changed by 
interaction with UV photons, atoms, radicals, ice mantles, and so on, 
when exposed to various environments in the ISM. Nano-diamonds are 
therefore likely to show different spectral features depending on 
their astrophysical environment. It is also possible that the surface 
was changed during the incorporation in the meteorite, during solar 
system formation.  However, it was shown by 
Virag et al.\, (\cite{virag+etal89}) that the presolar diamonds also 
carry anomalous hydrogen with $^{1}$H/$^{2}$D = 5193 (terrestrial = 6667) 
implying that at least some of the surface groups are presolar. 

The presented IR spectrum as well as most published IR data of 
meteoritic nano-diamonds represents well oxidized nano-diamonds. 
The potential user of the data needs to critically judge if these 
IR bands or possibly bands originating from another kind of 
surface species are likely to be produced by nano-diamonds in the 
environment under investigation.

\section{Derivation of optical constants}

\subsection{EELS measurement}

Electron energy loss spectroscopy (EELS) has been used before by 
Bernatowicz et al. (\cite{bernatowicz+etal90}) for the characterization 
of meteoritic nano-diamonds. It is performed by a magnetic prism 
spectrometer positioned beneath the viewing screen of a transmission 
electron microscope (TEM). During penetration of a thin sample, part of the 
electrons of the primary beam lose an energy $\Delta$E due to inelastic scattering 
at valence or core electrons of the sample material. The probability for 
such processes is proportional to the so-called loss function 
$\Im(-1/\varepsilon$($\Delta$E,$\hbar$q)), where $\varepsilon$($\Delta$E,$\hbar$q) 
is the complex dielectric function of the material and $\Im$ denotes the 
imaginary part. The quantity $\hbar$q is the momentum transfer to the materials' 
electron system in the scattering process. 
Resonances of the loss function occur for energies where the real part 
of the dielectric function vanishes ($\Re(\varepsilon$)=0) and collective 
excitations of the valence electrons (plasmons) are possible. This ``plasmon loss'' 
as well as losses due to interband transitions occur at loss energies 
lower than 100~eV, whereas at higher energies core electrons can be 
excited to states above the Fermi level (``core loss''). 

$\varepsilon$($\Delta$E,$\hbar$q) corresponds to the optical dielectric 
function $\varepsilon = m^2$ ($m$ being the complex refractive index) 
at wavelength $\lambda$ = hc/$\Delta$E under the condition 
that q$\approx$0, which restricts the scattering angle to small values. 
For practical purposes, q $\le$ 0.1~\AA$^{-1}$ is sufficient 
(Pfl\"uger et al.\ \cite{pflueger_palik}), at larger angles, deviations 
in the positions and widths of the plasmon peaks have been measured for 
carbonaceous materials (Falke et al.\ \cite{falke+etal95}). This technique, 
thus, allows to derive the optical data of a material in the energy 
range of the strong electronic transitions up to 100~eV, corresponding to 
wavelengths down to 0.01~$\mu$m.

We have prepared the Allende nano-diamonds for EELS measurements by dispersing 
them in ethanol and transferring them onto a TEM copper grid covered with a 
porous (``lacey'') carbon film. Measurements have been done exclusively on 
particle aggregates located over film holes. The EELS spectra were obtained 
with a Philips CM200FEG high-resolution TEM operated in the diffraction mode 
(camera length 400 - 4700~mm), using a Gatan Imaging Filter (GIF 100) with 
0.6~mm entrance aperture. 
Primary electron energies were 200 keV and the energy resolution was 0.9~eV 
based on the full-width-half-maximum (FWHM) of the zero energy loss peak. 
The acceptance angle for scattered electrons in the measurements was 0.4~mrad 
corresponding to a maximum momentum transfer of q $\approx$ 0.1~\AA$^{-1}$. 
The spectra were corrected for the influence of the energy distribution 
of the primary beam as well as for plural scattering events following the 
procedures described in Egerton (\cite{egerton96}). The contribution of the primary 
and elastically scattered electrons (``zero loss'') was removed from the 
spectrum by assuming a symmetric energy distribution of these electrons. 

\begin{figure}
\includegraphics[width=8.5cm,angle=0,clip]{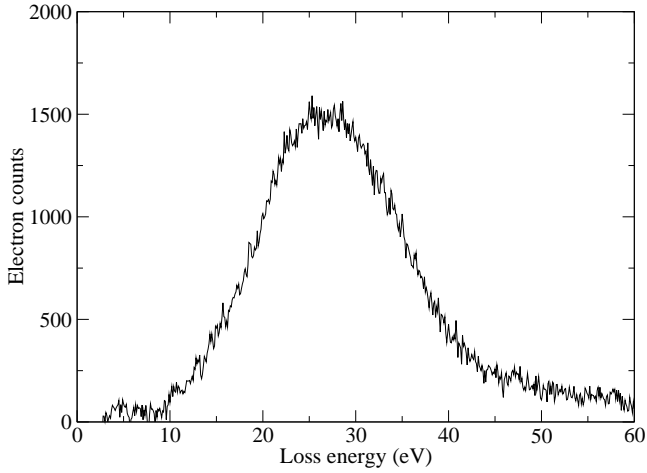}
\caption{A typical EELS spectrum of the Allende nano-diamonds. 
The maximum of the $\sigma$ plasmon (26.5~eV) is displaced 
compared to bulk diamonds (33.5~eV). 
This is due to a contribution from carbon atoms of a lower state of
hybridization resulting in a lower electron density.}
\label{EELSspectrum}
\end{figure}

A typical corrected energy loss spectrum of the meteoritic nano-diamonds 
is shown in Fig.\,\ref{EELSspectrum}. It is similar to the ones obtained 
by Bernatowicz et al. (\cite{bernatowicz+etal90}) but the plasmon peak 
is found at a smaller energy loss of 26.5~eV compared to about 29~eV, 
which is a consequence of the restriction of the momentum transfer. 
Compared to bulk diamond (peak at 33.5~eV), the feature is strongly 
shifted and broadened. 
According to Bernatowicz et al. (\cite{bernatowicz+etal90}), this can be 
understood as due to surface carbon atoms deviating in their bonding type 
from the volume atoms and a mixing of the corresponding dielectric functions. 
At the signal-to-noise level achievable with these angle-resolved EELS 
measurements, it is not possible to detect or exclude features due to 
$\pi$-electron transitions, which would occur at energies below 8~eV. 
In measurements performed with a much larger acceptance angle 
(in imaging mode), however, distinct $\pi$-electron features are 
typically not observed with our sample. For the determination of 
the dielectric function, the loss function at energy losses lower than 
8~eV has not been used. 

From the loss function Im(-1/$\varepsilon$($\Delta$E,$\hbar$q$\approx$0)), 
the dielectric function $\varepsilon$($\lambda$) can be determined via 
Kramers-Kronig analysis, if the spectrum can be normalized in a proper way. 
The possibility for the normalization is e.g. provided by assuming 
$\Re(\varepsilon)$ at low energies, the inverse value of which is related 
to the integral over the loss function weighted with $\lambda$ 
(Pfl\"uger et al.\ \cite{pflueger_palik}). 
The value of $\Re(\varepsilon)$ at low energies is identical to the 
square of the refractive index and can be determined from optical 
measurements in the visible. We adopted for $\Re(\varepsilon)$ a value 
of 4.0, estimated by Lewis et al.\ (\cite{lewis+etal89}) from density 
measurements for the meteoritic nano-diamonds. From the normalized 
loss function, $\varepsilon$($\lambda$) has been calculated in the 
wavelength range 10 - 200~nm. Further, this preliminary dataset has 
been converted into a synthetic absorption spectrum of nano-diamonds 
in an environment with $\varepsilon_m = 2.3$, which completes 
the optical absorption spectrum as shown in Fig. \ref{fullspec}.

\subsection{Optical constants}

\begin{figure}
\includegraphics[width=8.5cm,angle=0,clip]{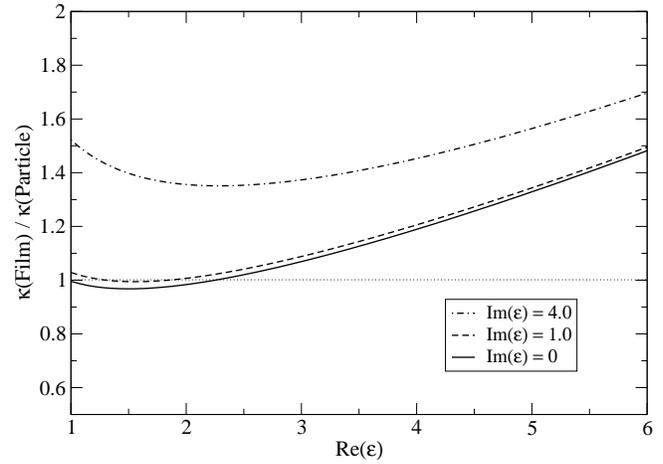}
\caption{Simulated ratio of thin-film and small-particle absorption 
depending on the dielectric function of the material.}
\label{film_particle}
\end{figure}

Before the combined absorption spectrum, shown in Fig.\, \ref{fullspec}, can be 
used to derive a consistent wavelength-dependent dielectric function, we have 
to ask whether the thin-film and small-particle absorption spectra are compatible 
or can be transformed into each other. Although they  match quite well in our 
measurements, the quantities $\kappa_{\rm film}$ and $\kappa_{\rm particle}$ 
(for spherical grains which are small compared to the wavelength) are related to 
the dielectric function $\varepsilon$ of the material by different expressions: 
\begin{equation}
\kappa_{\rm film} \, \rho \, = \, \frac{4 \, \pi \, k}{\lambda} \, = \, \frac{4 \, \pi}{\sqrt{2} \, \lambda} \, \sqrt{| \, \varepsilon \, | \, - \, \Re(\varepsilon)} \, .
\end{equation}
\begin{equation}
\kappa_{\rm particle} \, \rho \, = \, \frac{6 \, \pi \, \sqrt{\varepsilon_{\rm m}}}{\lambda} \, \Im \, \frac{\varepsilon - \varepsilon_{\rm m}}{\varepsilon + 2 \varepsilon_{\rm m}}
\label{kappa}
\end{equation}
with $\varepsilon_{\rm m}$ describing the surrounding medium (Bohren \& Huffman \cite{bohren+huffman83}). 
The ratio of $\kappa_{\rm film}$ and $\kappa_{\rm particle}$ for a range of 
complex $\varepsilon$-values and for $\varepsilon_{\rm m}$ = 2.3, 
which is approximately valid for both KBr and polyethylene matrices, 
is shown in Fig.\, \ref{film_particle}. It demonstrates that the deviation 
between both quantities for nano-diamonds with an assumed $\Re(\varepsilon)$ of 4.0 
is of the order of 20\,\% in spectral ranges where the absorption of the material 
is low ($\Im(\varepsilon) <$ 1.0). In the band edge, where both $\Re(\varepsilon)$ 
and $\Im(\varepsilon)$ increase, the deviation can reach 50\,\% or more, whereas 
at even shorter wavelengths, where anomalous dispersion occurs and $\Re(\varepsilon)$ 
falls down to values of 1.0 and smaller, the deviation is reduced again. This 
means that at wavelengths smaller than about 150~nm the steepness of the thin-film 
spectrum should decrease, what indeed seems to be observed (see Fig.\, \ref{fullspec}).

We have accounted for these effects by dividing the thin-film spectra by a factor 
of 1.3 and by not using them below $\lambda$ = 0.18~$\mu$m. After that, the 
different parts of the spectrum including the synthetic far-UV spectrum have 
been merged into one (using linear interpolation if necessary) at a reduced 
resolution, with equidistant points on the logarithmic scale. From this spectrum, 
the dielectric function has been deduced by using the Kramers-Kronig transformation 
based on Eq.~\ref{kappa} developed by Ossenkopf et al.\ (\cite{ossenkopf+etal92}). 

\begin{figure}
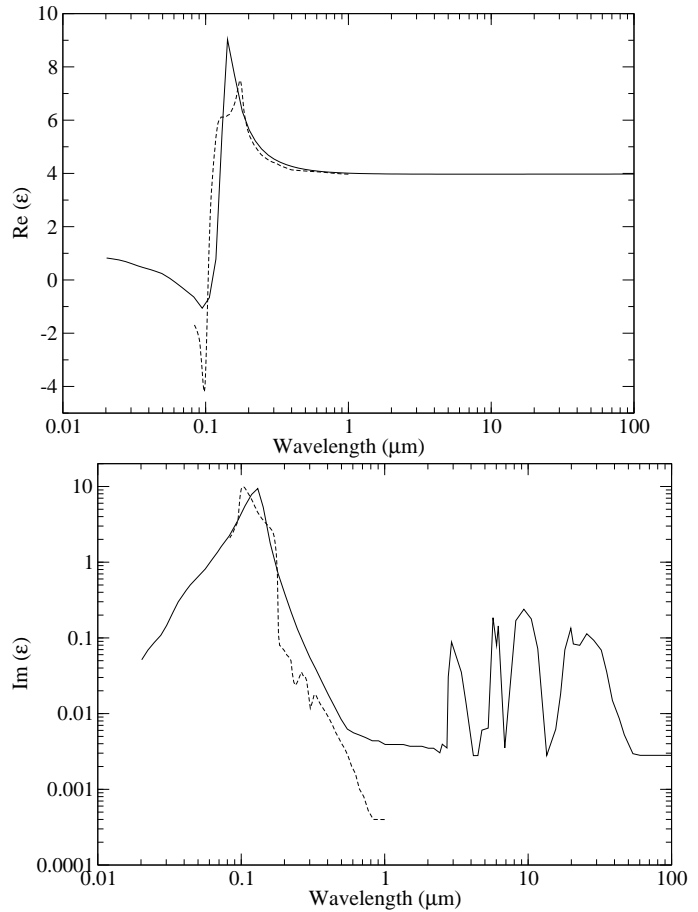

\includegraphics[width=8.5cm,angle=0,clip]{0544f6a.eps}
\includegraphics[width=9cm,angle=0,clip]{0544f6b.eps}
\caption{Dielectric function (upper panel: real part, lower panel: imaginary part) 
of the meteoritic nano-diamonds (solid line) compared to the previous diamond 
data (dashed line) published by Lewis et al.\, (\cite{lewis+etal89}).}
\label{df}
\end{figure}

The resulting dataset is shown in Fig.\, \ref{df} compared to the previous data 
by Lewis et al.\, (\cite{lewis+etal89}). In the UV and visible, the Lewis et al. 
data deviate from our new results. They are based on modified bulk data rather than 
on measurements and do not take into account the nano-particle character of the 
meteoritic diamonds. In astronomical modelling, usually the complex refractive 
index m = n + ik = $\sqrt{\varepsilon}$ is used instead of the dielectric function 
to describe the optical material properties. Therefore, we give these values 
in Table~\ref{n&k}. 

\section{Discussion}

\begin{figure*}
\includegraphics[width=15cm,angle=0,clip]{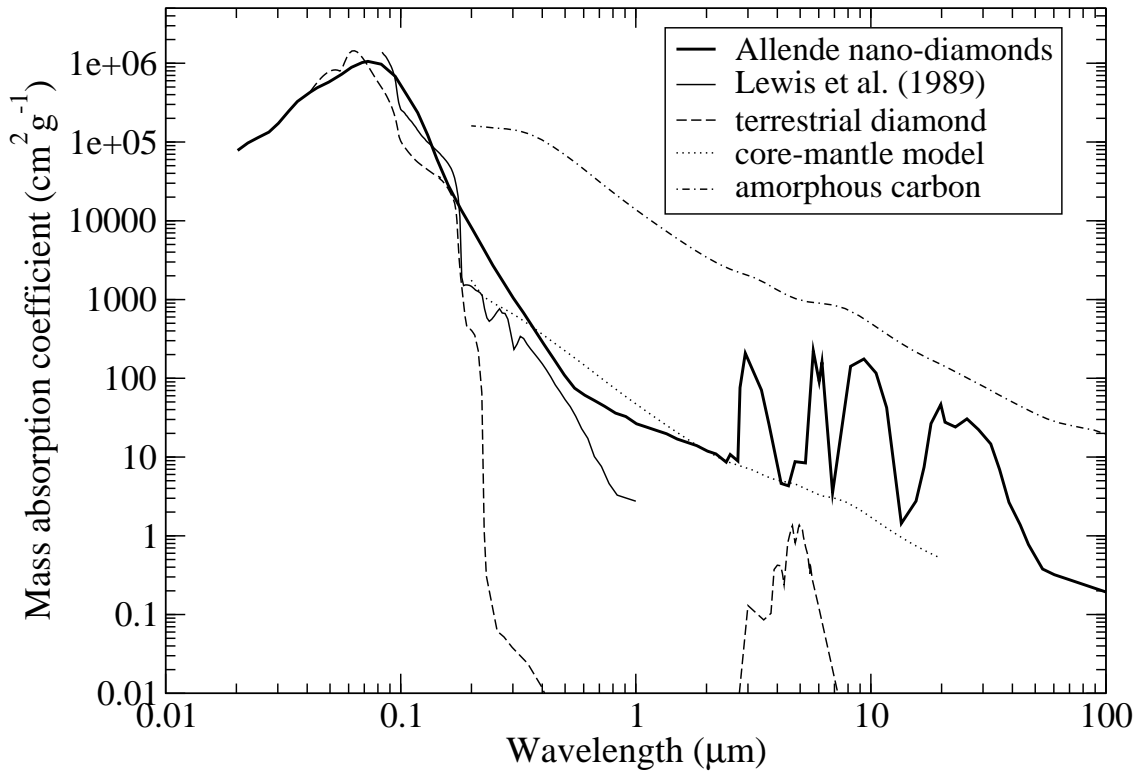}
\caption{Mass absorption coefficient spectra of the meteoritic 
nano-diamonds calculated from our data and from the data of 
Lewis et al.\ (\cite{lewis+etal89}). For comparison, the simulated 
absorption spectra of terrestrial diamond particles (data from 
Edwards and Philipp \cite{edwards+philipp85}), amorphous carbon 
particles (data "cel400" from J\"ager et al. \cite{jaeger+etal98}) 
and of a core-mantle particle with terrestrial diamond core 
(99\% volume fraction) and amorphous carbon mantle (1\% volume fraction) 
are shown. Using a more graphitic mantle material such as "cel1000" 
(not shown), results in a somewhat flatter IR continuum slope.}
\label{mac}
\end{figure*}

\begin{table*}
\caption{The optical constants $n$ and $k$ of meteoritic nano-diamond, 
$\lambda$ is the wavelength, $Q/a$ is the extinction efficiency.} 
\begin{center}
\begin{tabular}{|c|c|c|c||c|c|c|c||c|c|c|c|} \hline
$\lambda$ & Q/a & $n$ & $k$ & $\lambda$ & $Q/a$ & $n$ & $k$ & $\lambda$ & $Q/a$ & $n$ & $k$ \\ 
$\mu$m & cm$^{-1}$ &  & & $\mu$m & cm$^{-1}$ & & & $\mu$m & cm$^{-1}$ & & \\ \hline
0.0202 & 2.399E5 & 0.908 & 0.0283 & 0.4493 & 531.6 & 2.049 & 0.0030 & 6.1957 & 490.4 & 1.993 & 0.0361\\
0.0224 & 2.984E5 & 0.890 & 0.0388 & 0.4966 & 338.4 & 2.038 & 0.0021 & 6.8731 & 10.82 & 1.993 & 0.0008\\
0.0247 & 3.490E5 & 0.868 & 0.0499 & 0.5488 & 229.5 & 2.029 & 0.0015 & 8.2001 & 434.9 & 1.993 & 0.0423\\
0.0275 & 4.093E5 & 0.833 & 0.0649 & 0.6065 & 187.2 & 2.022 & 0.0014 & 9.3392 & 539.8 & 1.994 & 0.0599\\
0.0302 & 5.297E5 & 0.791 & 0.0921 & 0.6679 & 159.6 & 2.016 & 0.0013 & 10.521 & 357.5 & 1.994 & 0.0447\\
0.0334 & 7.601E5 & 0.743 & 0.1463 & 0.7361 & 135.2 & 2.012 & 0.0012 & 11.673 & 130.0 & 1.993 & 0.0180\\
0.0364 & 1.002E6 & 0.716 & 0.2088 & 0.8198 & 110.6 & 2.008 & 0.0011 & 13.478 & 4.401 & 1.993 & 0.0007\\
0.0413 & 1.318E6 & 0.691 & 0.3061 & 0.9067 & 100.3 & 2.005 & 0.0011 & 15.567 & 8.455 & 1.993 & 0.0016\\
0.0442 & 1.500E6 & 0.684 & 0.3660 & 1.0063 & 81.14 & 2.002 & 0.0010 & 16.836 & 22.83 & 1.993 & 0.0046\\
0.0494 & 1.767E6 & 0.676 & 0.4642 & 1.1052 & 73.99 & 2.001 & 0.0010 & 18.026 & 81.64 & 1.993 & 0.0175\\
0.0561 & 2.226E6 & 0.661 & 0.6143 & 1.2140 & 67.44 & 1.999 & 0.0010 & 19.834 & 142.5 & 1.994 & 0.0336\\
0.0617 & 2.719E6 & 0.688 & 0.7589 & 1.3521 & 60.62 & 1.998 & 0.0010 & 20.666 & 84.96 & 1.994 & 0.0209\\
0.0685 & 3.131E6 & 0.744 & 0.9170 & 1.5006 & 51.88 & 1.997 & 0.0009 & 22.878 & 73.73 & 1.994 & 0.0200\\
0.0725 & 3.239E6 & 0.793 & 1.0123 & 1.6482 & 47.26 & 1.996 & 0.0009 & 25.624 & 93.56 & 1.994 & 0.0285\\
0.0826 & 2.981E6 & 0.917 & 1.2159 & 1.8166 & 42.90 & 1.995 & 0.0009 & 28.680 & 68.44 & 1.994 & 0.0233\\
0.0946 & 2.099E6 & 1.149 & 1.5429 & 2.0162 & 36.47 & 1.995 & 0.0009 & 32.369 & 44.98 & 1.994 & 0.0173\\
0.1062 & 1.215E6 & 1.564 & 1.7640 & 2.1915 & 33.56 & 1.994 & 0.0009 & 35.291 & 21.32 & 1.994 & 0.0089\\
0.1181 & 7.240E5 & 2.080 & 1.8828 & 2.4239 & 26.40 & 1.994 & 0.0008 & 38.699 & 8.142 & 1.994 & 0.0037\\
0.1303 & 3.738E5 & 2.870 & 0.6377 & 2.5208 & 33.03 & 1.994 & 0.0010 & 43.177 & 4.220 & 1.994 & 0.0022\\
0.1423 & 1.882E5 & 3.124 & 0.8535 & 2.7167 & 27.36 & 1.994 & 0.0009 & 46.804 & 2.373 & 1.994 & 0.0013\\
0.1592 & 8.657E4 & 2.793 & 0.3182 & 2.7753 & 233.4 & 1.994 & 0.0077 & 53.746 & 1.161 & 1.994 & 0.0007\\
0.1807 & 4.226E4 & 2.520 & 0.1404 & 2.9240 & 638.1 & 1.994 & 0.0222 & 60.340 & 0.983 & 1.994 & 0.0007\\
0.2019 & 2.399E4 & 2.374 & 0.0790 & 3.4233 & 217.2 & 1.994 & 0.0088 & 66.686 & 0.889 & 1.994 & 0.0007\\
0.2231 & 1.411E4 & 2.285 & 0.0477 & 3.6873 & 79.40 & 1.993 & 0.0035 & 73.700 & 0.804 & 1.994 & 0.0007\\
0.2466 &    8320 & 2.218 & 0.0294 & 4.1531 & 14.17 & 1.993 & 0.0007 & 81.450 & 0.728 & 1.994 & 0.0007\\
0.2747 &    4984 & 2.165 & 0.0187 & 4.4645 & 13.21 & 1.993 & 0.0007 & 90.017 & 0.659 & 1.994 & 0.0007\\
0.3012 &    3209 & 2.131 & 0.0129 & 4.7597 & 26.83 & 1.993 & 0.0015 & 98.620 & 0.601 & 1.994 & 0.0007\\
0.3329 &    2084 & 2.102 & 0.0090 & 5.2659 & 25.81 & 1.993 & 0.0016 & 109.95 & 0.539 & 1.994 & 0.0007\\
0.3679 &    1312 & 2.080 & 0.0062 & 5.6899 & 685.1 & 1.994 & 0.0460 &  &  &  & \\
0.4066 &   827.2 & 2.063 & 0.0042 & 6.0254 & 279.1 & 1.993 & 0.0200 &  &  &  & \\ \hline
\end{tabular}
\end{center}
\label{n&k}
\end{table*}

Observations of nano-diamonds in space would be very desirable first for 
the clarification of their origin and evolution, and second for the 
determination of their abundance. Unfortunately, with the exception of 
the diamond detection in a few Herbig Ae/Be objects and one post-AGB star 
(Guillois et al.\ \cite{guillois+etal99}), there were no positive results yet. 
There might be a number of reasons for this; one is that the intrinsic 
features longward of the electronic band edge are very weak, even for 
nitrogen-doped diamonds. A second reason is that surface groups may 
produce strong bands, but that it may be hard to distinguish them 
from bands produced by other carbonaceous matter. 
Moreover, the actual surface structure of nano-diamonds in space is 
unknown and is likely to depend on the environment. Finally, the 
diamonds may be covered or incorporated into larger grains, which would 
alter their signature or prevent it from becoming visible. 

One chance for the detection is the tiny size of the nano-diamonds, which 
opens the possibility of stochastic heating to temporarily high temperatures 
by single UV-photon absorption. Consequently, they could become visible in 
emission, if the nano-diamonds would be separated from other grains. 
In the case of the diamond emission bands seen in the Herbig Ae/Be objects, 
it has been possible to derive an abundance of the diamonds which, however, 
is very low (1 part per billion relative to hydrogen, van Kerckhoven et al.\ 
\cite{vankerckhoven+etal02}). This abundance value, however, may reflect 
only a population of diamond grains in a narrow size range which allows 
both the stochastic heating process and the presence of 
well-ordered surfaces. The abundance of nano-diamonds in primitive 
meteorites is, as we know, much higher (500--1000~ppm), but one 
must bear in mind the possibility that
the concentration of interstellar diamonds in meteorites could be
disproportionately high as a result of their superior chemical and
physical durability (Sandford \cite{sandford96}).

A more exact way to derive upper limits of the diamond abundance 
in space is provided by the ultraviolet absorption of the diamonds. 
Lewis et al.\ (\cite{lewis+etal89}) suggested that it would be 
compatible with the measured interstellar extinction 
(e.g.\ Stecher \cite{stecher65}; Fitzpatrick \& Massa 
\cite{fitzpatrick+massa86}; Clayton et al.\, \cite{clayton+etal03}), 
if $<$10\% of the carbon would be present as diamonds. 
Interestingly, the comparison of mass absorption coefficient spectra 
in Fig.\, \ref{mac} reveals that the Lewis et al. data are erroneous 
because they produce higher absorption values than the 
original bulk diamond data, contrary to what had been intended. 
The reason for this is that Lewis et al.\ (\cite{lewis+etal89}) 
reduced the imaginary part of the dielectric function to account 
for the reduced density of the meteoritic diamonds instead of reducing 
the polarizability. Our results give a value of the ultraviolet 
extinction of meteoritic nano-diamonds at the Lyman limit which 
is about a factor of 1.5 less than the estimate by 
Lewis et al.\ (\cite{lewis+etal89}). Thus, the restrictions on 
the amounts of diamonds which can be present in space are less 
strict than previously suggested. Even larger masses of big diamonds 
could be hidden in the ISM, in principle.  

Another constraint on the abundance of nano-diamonds comes from the 
lack of structure in the observed galactic extinction curve at the 
position of the diamond band edge (6.9~eV). 
Lewis et al.\ (\cite{lewis+etal89}) estimated a detection threshold of 
this structure of about 10\% of the carbon present as diamonds. 
This threshold should be increased as well to account for the shallower 
band edge of the meteoritic nano-diamonds compared to the data used 
by Lewis et al.\ (\cite{lewis+etal89}). 

For the energetic and hydrodynamic modelling of environments 
possibly containing significant amounts of diamond dust, it will be 
important to consider the infrared properties of these particles. 
Our results indicate that their mass absorption coefficient in the 
visible and IR is about two orders of magnitude lower than that of 
amorphous carbon particles. That would not have been expected from 
previous results e.g. by Bernatowicz et al.\ (\cite{bernatowicz+etal90}), 
which indicated the presence of a thick amorphous-carbon mantle of 
up to 46\% of the particle's volume. We find that a core-mantle model
is probably not well suited for the description of meteoritic nano-diamonds 
because our spectroscopic results would be consistent only with an 
unphysical mantle volume fraction of not more than 1\% (see Fig.\, \ref{mac}).

\section{Conclusions}

We have performed an extended spectroscopic study of presolar nano-diamonds 
from the Allende meteorite to close the gaps in the currently known optical data 
of this material. Different spectroscopic techniques have been used to obtain 
a consistent {\it complete absorption spectrum} of our sample of 
meteoritic nano-diamonds and to derive the optical constants. Our results 
indicate that the limits on the amount of diamond grains which can be present 
in space within the current observational restraints are about a factor of 
1.5 higher than previously estimated by Lewis et al.\ (\cite{lewis+etal89}). 
Furthermore, the mass absorption coefficient at visible and IR wavelengths 
is about two orders of magnitude lower than that of amorphous carbon, which 
is not consistent with the presence of a thick amorphous-carbon mantle. 
These results can be used to improve current models of the environments 
where nano-diamonds are expected to be present. Further studies on 
the optical properties of nano-diamonds in a wide spectral range would be 
valuable, in order to investigate their dependence on surface processing 
in the lab and in space.

\begin{acknowledgements}
      The authors would like to thank Gabriele Born for help with the preparational 
      work. Further, we thank R. Schl\"ogl, D. Su, H. Sauer, M. Willinger, and 
      J.-O. M\"uller of the Fritz-Haber-Institut Berlin 
      for the possibility to perform the EELS measurements and their help. This work 
      has been supported by DFG grant Mu~1164/4 within the DFG Research Unit 
      "Laboratory Astrophysics". It is also part of a joint effort in laboratory 
      astrophysics between the University of Jena and the MPI for Astronomy Heidelberg.
      ACA acknowledges support from the Carlsberg Foundation. 
\end{acknowledgements}


\begin{thebibliography}{}
\bibitem[]{} 
\bibitem[1990]{alexander+etal90} Alexander C.\,M.\,O'D., Swan P. \& 
  Walker R.\,M., 1990, Nat 348, 715
%\bibitem[1990]{amari+etal90} Amari S., Anders E., Virag A. \& Zinner E.\,K., 1990, Nat 345, 238
\bibitem[1994]{amari+etal94} Amari S., Lewis R.S. \& Anders E., 1994, Geochim.\ Cosmochim.\ Acta 58, 459
%\bibitem[1989]{anders+grevesse89} Anders E. \& Grevesse N., 1989, Geochim.\ Cosmochim.\ Acta 53, 197
%\bibitem[1993]{anders+zinner93} Anders E. \& Zinner E.\,K., 1993, Meteoritics 28, 490 
\bibitem[1998]{andersen+etal98} Andersen A.C., J{\o}rgensen U.G., Nicolaisen F., 
  S{\o}rensen P.G. \& Glejb{\o}l K., 1998, A\&A 330, 1080
\bibitem[1992]{allamandola+etal92} Allamandola L.J., Sandford S.A. \& Tielens A.G.G.M., 1992, A\&A 399, 134
\bibitem[1998]{banhart+etal98} Banhart F., Lyatovich Y., Braatz A., et al., 1998, 
  Meteorit.\ Planet.\ Sci.\ 33, A12
%\bibitem[1987]{bernatowicz+etal87} Bernatowicz T.\,J., Fraundorf G., Tang M., et al., 1987, Nat 330, 728
%\bibitem[1991]{bernatowicz+etal91} Bernatowicz T.\,J., Amari S., Zinner E.\,K.  \& Lewis R.\,S., 1991, ApJ 373, L73
\bibitem[1990]{bernatowicz+etal90} Bernatowicz T.J., Gibbons P.C. \& Lewis R.S., 1990, ApJ 359, 246
%\bibitem[1999]{besmehn+etal99} Besmehn A., Hoppe P., Strebel R. \& Ott U., 1999, Meteorit.\ Planet.\ Sci.\ 34, A11
\bibitem[1988]{blake+etal88} Blake D.F., Freund F., Krishnan K.F.M., et al., 1988 
  Nature 332, 611
\bibitem[1983]{bohren+huffman83} Bohren C.F. \& Huffman D.R., 1983, Absorption and Scattering of Light by Small Particles (John Wiley \& Sons, New York)
\bibitem[2000]{braatzthesis} Braatz A., 2000, PhD thesis, Friedrich Schiller University Jena
\bibitem[2000]{braatz+etal00} Braatz A., Ott U., Henning Th., J\"ager C., Jeschke G., 2000, 
  Meteorit.\ Planet.\ Sci.\ 35, 75
\bibitem[1996]{brooke+etal96} Brooke T.Y., Sellgren K. \& Smith R.G., 1996, ApJ 459, 209
\bibitem[1999]{brooke+etal99} Brooke T.Y., Sellgren K. \& Geballe T.R., 1999, ApJ 517, 883
%\bibitem[1990]{buss+etal90} Buss R.H.Jr., Cohen M., Tielens A.G.G.M., et al., 1990, ApJ 365, 23
%\bibitem[1987]{bussoletti87}Bussoletti E., Colangeli L., Borghesi A. \& Orofino V., 1987, A\&AS 70, 257
%\bibitem[1997]{cheng+etal97} Cheng C.-L., Chang H.-C. \& Lin J.-C., 1997, J.\ Chem.\ Phys.\ 106, 7411 {\it is this the right reference?????}
\bibitem[1989]{clayton89} Clayton D.D., 1989, ApJ 340, 613
\bibitem[1995]{clayton+etal95} Clayton D.D., Mayer B.S., Sanderson C.I., et al., 1995, ApJ 447, 894
\bibitem[2003]{clayton+etal03} Clayton G.C., Gordon K.D., Salama F., et al., 2003, ApJ 592, 947
\bibitem[1994]{colangeli+etal94} Colangeli L., Mennella V., Stephens J.R. \& Bussoletti E., 1994, A\&A 284, 583
%\bibitem[1995]{colangeli+etal95}Colangeli L., Menella V., Palumbo P., et al., 1995, A\&AS 113, 561
%\bibitem[1994]{daulton94} Daulton T.L., Eisenhour D.D., Lewis R.S \& Bernatowicz T.J., 1994, Lunar Planet.\ Sci.\ 25, 313
\bibitem[2002]{dai+etal02} Dai Z.R., Bradley J.P., Joswiak D.J., et al., 2002, Nat 418, 157
\bibitem[1970]{daniels+etal70} Daniels J., v. Festenberg, C., Raether H., \& Zeppenfeld K., 1970, 
  Springer Tracts in Modern Physics, 94, 77
%\bibitem[1996]{daulton+etal96} Daulton T.\,L., Eisenhour D.\,D., Bernatowitz T.\,J., et al., 1996, Geochim.\ Cosmochim.\ Acta 60, 4853
%\bibitem[1977]{davies77} Davies G., 1977, The optical properties of diamond. In: P.\,L.\,Walker, P.\,A\,Thrower (eds.) Chemistry and Physics of Carbon 13, 2
%\bibitem[1909]{debye09} Debye P., 1909, Ann.\ Phys.\ 30, 57
\bibitem[1996]{dorschner+etal96} Dorschner J., Henning Th., J\"ager C.,
  Mutschke H. \& Ott U., 1996, Meteorit. Planet. Sci. 31, A37
\bibitem[1985]{edwards+philipp85} Edwards D.F. \& Philipp H.R., 1985, In: E.\,D.\,Palik (ed.), 
Handbook of Optical Constants of Solids, Orlando: Academic Press, p. 665
\bibitem[1996]{egerton96} Egerton R.F., 1996, Electron Energy Loss Spectroscopy in the Electron Microscope, 
  2nd ed., Plenum Press, New York
\bibitem[1992]{evans92} Evans S., 1992, Surface properties of diamond. 
  In: J.E.\,Field (ed.), The Properties of Natural and Synthetic Diamond, 
  Academic Press, London, 181
\bibitem[1994]{evans94} Evans S., 1994, Reactivity of diamond surfaces. 
  In: G.\,Davies (ed.), Properties and Growth of Diamond, 
  EMIS Datareviews Series, vol.9, INSPEC, London, 64
%\bibitem[1997]{evans97} Evans T., 1997, Aggregation of nitrogen in diamond. In: J.\,Field (ed.) The Properties of Natural and Synthetic Diamonds, Academic Press, Harcourt Brace \& Company, 259
\bibitem[1995]{falke+etal95} Falke U., Weber, A. \& Ullmann J., 1995, 
  Microsc. Microanal. Microstruct., 6, 113
\bibitem[1986]{fitzpatrick+massa86} Fitzpatrick E.L. \& Massa D., 1986, ApJ 307, 286
\bibitem[1989]{fraundorf+etal89} Fraundorf P., Fraundorf G., Bernatowitz T., et al.,
  1989, Ultramicroscopy 27, 401
\bibitem[1999]{guillois+etal99} Guillois O., Ledoux G. \& Reynaud C., 1999, ApJ 521, L133
\bibitem[1997]{hill+etal97} Hill H.G.M., d'Hendecourt L.B., Perron C. \& Jones A.\,P., 1997, 
  Meteoritics \& Planetary Sci.\ 32, 713
\bibitem[1998]{hill+etal98} Hill H.G.M., Jones A.\,P. \& d'Hendecourt L.B., 1998, A\&A 336, L41
%\bibitem[1993]{hoyle+wichramasinghe93} Hoyle F. \& Wichramasinghe N.\,C., 1993, Ap\&SS 207, 309
\bibitem[1990]{huss90} Huss G.\,R., 1990, Nat 347, 159
%\bibitem[1997]{huss97} Huss G.R., 1997, The survival of presolar grains in solar system bodies.  In: T.J.\,Bernatowicz \& E.\,Zinner (eds.) Astrophysical Implications of the Laboratory Study of Presolar Materials, AIP 402, AIP, 721
%\bibitem[1997]{huss+etal97} Huss G.\,R., Hutcheon I.\,C. \& Wasserburg G.\,J., 1997, Geochim.\ Cosmochim.\ Acta 61, 5117
\bibitem[1995]{huss+lewis95} Huss G. \& Lewis R.S., 1995, Geochim.\ Cosmochim.\ Acta 52, 115
%\bibitem[1994]{hutcheon+etal94} Hutcheon I.\,D., Huss G.\,R., Fahey A.\,J. \& Wasserburg G.\,J., 1994, ApJ 425, L97
\bibitem[1997]{connythesis} J\"ager C., 1997, PhD thesis, Friedrich Schiller University Jena
\bibitem[1998]{jaeger+etal98} J\"ager C., Mutschke H., Henning Th., 
  1998, A\&A 332, 291
\bibitem[2000]{jones+dhendecourt00} Jones A.P. \& d'Hendecourt L., 2000, 
  A\&A 355, 1191
\bibitem[2004]{jones+etal04} Jones A.P., d'Hendecourt L.B., Sheu S.-Y., Chang H.-C., Cheng
C.-L., Hill H.G.M., 2004, A\&A 416, 235
  A\&A 355, 1191
\bibitem[1988]{jorgensen88} J{\o}rgensen U.G., 1988, Nat 332, 702
\bibitem[1999]{jorgensen+andersen99} J{\o}rgensen U.\,G. \& Andersen A.\,C., 1999, 
  In: R.\,F.\,Wing (ed.) The Carbon Star Phenomenon, IAU Symp.\ 177, Kluwer, 349
%\bibitem[1996]{kehm+etal96} Kehm K., Amari S., Hohenberg C.\,M. \& Lewis R.\,S., 1996, Lunar Planet.\ Sci.\ 27, 657
\bibitem[1995]{koike+etal95} Koike C., Wickramasinghe C., Kano N., et al., 1995, MNRAS 277, 986
\bibitem[1989]{kwok+etal89} Kwok S., Volk K.\,M. \& Hrivnak B.\,J., 1989, ApJ 345, L51
\bibitem[1989]{kwok+etal99} Kwok S., Volk K.\,M. \& Hrivnak B.\,J., 1999, 
in Asymptotic Giant Branch Stars, IAU Symposium 191, ed. T. Le Bertre, 
A. Lebre, \& C. Waelkens, 297
\bibitem[1992]{lambert92} Lambert D.\,L., 1992, A\&AR 3, 201
%\bibitem[1988]{lee88} Lee T., 1988, Implications of isotopic anomalies on nucleosynthesis. In: J.\,F.\,Kerridge , M.\,S.\,Matthews (eds.) Meteorites and the Early Solar System, University of Arizona Press, Tucson Arizona, 1063
\bibitem[1987]{lewis+etal87} Lewis R.\,S., Tang M., Wacker J.\,F., et al., 1987, Nat 326, 160
\bibitem[1989]{lewis+etal89} Lewis R.S., Anders E. \& Draine B.T., 1989, Nat 339, 117
%\bibitem[1990]{lewis+etal90} Lewis R.\,S., Amari S. \& Anders E., 1990, Nat 348, 293
\bibitem[1991]{lewis+etal91} Lewis R.\,S., Huss G.\,R. \& Lugmair G., 1991, Lunar Planet.\ Sci.\ 22, 807
%\bibitem[1994]{lewis+etal94} Lewis R.\,S., Amari S. \& Anders E., 1994, Geochim.\ Cosmochim.\ Acta 58, 471
%\bibitem[1890]{lorenz90} Lorenz L., 1890, Vidensk. Selsk. Skr. T. VI(6), (Bianco Lunos Kgl. Hof-Bogtrykkeri, Copenhagen), 1
%\bibitem[1908]{mie08} Mie, G., 1908, Ann.\ Phys.\ Leipz.\ 25, 377
\bibitem[1995]{mutschke+etal95} Mutschke H., Dorschner J., Henning Th., et al., 1995, ApJ 454, L157
%\bibitem[1991]{nichols+etal91} Nichols R.\,H.\,Jr., Hohenberg C.\,M., Amari S. \& Lewis R.\,S., 1991, Meteoritics 26, 377
%\bibitem[1994]{nichols+etal94} Nichols R.\,H.\,Jr., Kehm K., Brazzle R., et al., 1994, Meteoritics 29, 510
%\bibitem[1994]{nittler+etal94} Nittler L.\,R., Alexander C.\,M.\,O'D., Gao X., et al., 1994, Nat 370, 443 
%\bibitem[1995]{nittler+etal95} Nittler L.\,R., Hoppe P., Alexander C.\,M.\,O'D., et al., 1995, ApJ 453, L25 
%\bibitem[2003]{okada+etal03} Okada K., Kimoto K., Komatsu S. \& Matsumoto S., 2003, J. Appl. Phys. 93, 3120
%\bibitem[1991]{Ossenkopf} Ossenkopf V, 1991,  A\&A 251, 210
\bibitem[1992]{ossenkopf+etal92} Ossenkopf V., Henning Th. \& Mathis J.S., 1992, A\&A 261, 567
%\bibitem[1990]{ott+begemann90} Ott U., Begemann F., 1990, ApJ 353, L57
%\bibitem[1993]{ott93} Ott U., 1993, Nat 364, 25
\bibitem[1996]{ott96} Ott U., 1996, ApJ 463, 344
\bibitem[2003]{ott03} Ott U., 2003, In: Th. Henning (ed.) Astromineralogy, Springer, Berlin, 236
\bibitem[2000]{papoular00} Papoular R., 2000, A\&A 362, L9
\bibitem[1991]{pflueger_palik} Pfl\"uger J. \& Fink J., 1991, In: E.\,D.\,Palik (ed.), 
Handbook of Optical Constants of Solids II, Boston: Academic Press, p. 293
%\bibitem[1999]{phelps99} Phelps A.W., 1999, Lunar Planet.\ Sci.\ 30, 1749
\bibitem[2004]{posch+etal04} Posch Th., Mutschke H., Andersen A.C., ApJ, submitted
%\bibitem[1993]{prombo+etal93} Prombo C.\,A., Podosek F.\,A., Amari S., Lewis R.\,S., 1993, ApJ 410, 393
%\bibitem[1993]{richter+etal93} Richter S., Ott U., Begemann F., 1993, $s$-process isotope abundance anomalies in meteoritic silicon carbide: new data.  In: M.\,Busso, R.\,Gallino, C.\,M.\,Raiteri (eds.) Nuclei in the Cosmos III, AIP 327, AIP Conf.\ Ser.\, 127
\bibitem[1991]{russell+etal91} Russell S.S., Arden J.W., Phillinger C.T., 1991, Sci 254, 1188
%\bibitem[1993]{russell+etal93} Russell S.\,S., Alexander C.\,M.\,O'D., Ott U., Zinner E.\,K., Arden J.\,W., Pillinger C.\,T., 1993, Meteoritics 28, 425
%\bibitem[1995]{russell+etal95} Russell S.\,S., Lee M.\,R., Arden J.\,W., Pillinger C.\,T., 1995, Meteoritics 30, 399
%\bibitem[1996]{russell+etal96} Russell S.\,S., Arden J.\,W., Pillinger C.\,T., 1996, Meteorit.\ Planet.\ Sci.\ 31, 343
%\bibitem[1997]{russell+etal97} Russell S.\,S., Ott U., Alexander C.\,M.\,O'D., Zinner E., Arden J.\,W., Pillinger C.\,T., 1997, Meteoritics \& Plan.\ Sci. 32, 719
%\bibitem[1991]{sandford+etal91} Sandford S.A., Allamandola L.J., Tielens A.G.G.M., Sellgren K., Tapia M., Pendleton Y., 1991, ApJ 371, 607
\bibitem[1996]{sandford96} Sandford S.A., 1996, Meteorit.\ Planet.\ Sci.\ 31, 449
%\bibitem[1969]{saslaw+gustad69} Saslaw W.\,C., Gaustad J.\,E., 1969, Nat 221, 160
%\bibitem[1994]{sedlmayr94} Sedlmayr E., 1994, in Molecules in the Stellar Environment, LNP 428, ed.\ U.G. J{\o}rgensen (Springer, Berlin), 163
\bibitem[1980]{sekiyamamoto80} Seki J. \& Yamamoto T., 1980, Astrophys. Space Sci., 72, 79
\bibitem[2002]{sheu+etal02} Sheu S.-Y., Lee I.-P., Lee Y.T., \& Chang H.-C., 2002, ApJ 581, L55
\bibitem[2003]{speck+hofm03} Speck A.K. \& Hofmeister A.M., 2003, ApJ, in press
\bibitem[1965]{stecher65} Stecher T.P., 1965, ApJ 142, 1683
\bibitem[1988]{tang+anders88} Tang M. \& Anders E., 1988, Geochim.\ Cosmochim.\ Acta 52, 1235
\bibitem[1987]{tielens+etal87} Tielens A.G.G.M., Seab C.G., Hollenbach D.J., McKee C.F., 1987, 
ApJ 319, L109
\bibitem[2002]{vankerckhoven+etal02} van Kerckhoven C., Tielens A.G.G.M. \& Waelkens C., 2002 A\&A 384, 568
\bibitem[1989]{virag+etal89} Virag A., Zinner E.\,K.,  Lewis R.\,S., Tang M., 1989, 
  Lunar Planet.\ Sci.\ 20, 1158
%\bibitem[1992]{virag+etal92} Virag A., Wopenka B., Amari S., Zinner E.\,K., Anders E., Lewis R.\,S., 1992, Geochim. Cosmochim. Acta 56, 1715
\bibitem[1989]{volk+etal99} Volk K.\,M. Kwok S. \& Hrivnak B.\,J., 1999, ApJ 516, L99
\bibitem[2000]{vonhelden+etal00} von Helden G., Tielens A.G.G.M., van Heijnsbergen D., et al., 2000, 
  Science 288, 313
%\bibitem[1995]{wasserburg+etal95} Wasserburg G.\,J., Boothroyd A.\,I., Sackmann I.\,J., 1995, ApJ 447, 37
%\bibitem[1995]{zinner95} Zinner E.\,K. 1995, Interstellar grains from primitive meteorites: New constraints on nucleosynthesis theory and stellar evolution models. In: M.\,Busso, R.\,Gallino, C.\,M.\,Raiteri (eds.) Nuclei in the Cosmos III, AIP 327, 567
\bibitem[1998]{zinner98} Zinner E., 1998, Annu.\ Rev.\ Earth Planet.\ Sci.\ 26, 147

\end{thebibliography}
\end{document}